\documentclass{article}
\usepackage{placeins}

\usepackage{arxiv}

\usepackage[utf8]{inputenc} % allow utf-8 input
\usepackage[T1]{fontenc}    % use 8-bit T1 fonts
\usepackage{hyperref}       % hyperlinks
\usepackage{url}            % simple URL typesetting
\usepackage{amsfonts}       % blackboard math symbols
\usepackage{nicefrac}       % compact symbols for 1/2, etc.
\usepackage{microtype}      % microtypography
\usepackage{lipsum}
\usepackage{graphicx}
\graphicspath{ {./images/} }
\usepackage{graphicx}
\usepackage[onehalfspacing]{setspace}

\usepackage{orcidlink}

\usepackage{tocbasic}
\usepackage{booktabs}
\usepackage{multicol}
\usepackage{fontawesome5}
\usepackage{multirow}
\usepackage{algorithm}
\usepackage{algpseudocode}
\usepackage{caption}
\usepackage{tabularray}
\usepackage{subcaption}
\usepackage{array}

\usepackage{lscape} % For landscape tables (if needed)
\usepackage{ragged2e} % For better text alignment
\usepackage{longtable} % For tables spanning multiple pages

\title{Between Innovation and Oversight: A Cross-Regional Study of AI Risk Management Frameworks in the EU, U.S., UK, and China}

\author{
  Amir Al-Maamari \orcidlink{0009-0004-9816-2362} \\
  Faculty of Computer Science and Mathematics\\
  University of Passau\\
  Germany, Passau \\
  \texttt{almaam03@ads.uni-passau.de} \\
}

\begin{document}
\maketitle
\begin{abstract}
As artificial intelligence (AI) technologies increasingly enter important sectors like healthcare, transportation, and finance, the development of effective governance frameworks is crucial for dealing with ethical, security, and societal risks. This paper conducts a comparative analysis of AI risk management strategies across the European Union (EU), United States (U.S.), United Kingdom (UK), and China. A multi-method qualitative approach, including comparative policy analysis, thematic analysis, and case studies, investigates how these regions classify AI risks, implement compliance measures, structure oversight, prioritize transparency, and respond to emerging innovations. Examples from high-risk contexts like healthcare diagnostics, autonomous vehicles, fintech, and facial recognition demonstrate the advantages and limitations of different regulatory models. 
The findings show that the EU implements a structured, risk-based framework that prioritizes transparency and conformity assessments, while the U.S. uses decentralized, sector-specific regulations that promote innovation but may lead to fragmented enforcement. The flexible, sector-specific strategy of the UK facilitates agile responses but may lead to inconsistent coverage across domains. China's centralized directives allow rapid large-scale implementation while constraining public transparency and external oversight. These insights show the necessity for AI regulation that is globally informed yet context-sensitive, aiming to balance effective risk management with technological progress. The paper concludes with policy recommendations and suggestions for future research aimed at enhancing effective, adaptive, and inclusive AI governance globally.

\keywords{AI Governance, AI Risk Management Frameworks, EU AI Act, Comparative Analysis}
\end{abstract}

\section{Introduction}\label{sec:intro}
Artificial intelligence (AI) has transitioned from a specialized research domain to a critical technology influencing almost all global industries \cite{bostrom2014paths, floridi2022unified}. The applications include healthcare diagnostics, autonomous vehicles, financial analytics, and personalized consumer services \cite{RASHID2024100277}. As AI systems gain prevalence and capability, concerns about their ethical, security, and societal implications have intensified \cite{dignum2018ethics}. Concerns include potential biases in algorithmic decision-making, privacy erosion, and the broader social impacts of automation and surveillance technologies \cite{blobel2020autonomous, Pratt2022}.

Policymakers and regulatory bodies have begun to formulate governance strategies aimed at ensuring the responsible development and use of AI \cite{batool2023responsible}. The European Union (EU) initiated a significant legislative effort with its proposed AI Act, set to take effect on August 1, 2024, and to be fully implemented by August 1, 2027 \cite{eu2021proposal}. This regulation utilizes a risk-based categorization that enforces progressively stricter requirements on applications classified as high-risk, with the objective of establishing a global standard for AI oversight \cite{Almada2025}. Historically, the United States has employed a decentralized approach, with federal agencies like the Food and Drug Administration (FDA) and state governments establishing domain-specific guidelines. This has led to a fragmented yet innovation-friendly environment \cite{cath2018artificial, morley2023operationalising}. The United Kingdom (UK) adopts a sector-specific and flexible regulatory framework aimed at balancing business competitiveness with accountability \cite{lords2018ai}. In contrast, China's governance model is characterized by a more centralized and state-led approach, which aligns AI deployment with national priorities and utilizes government oversight to direct technological innovation \cite{ding2018deciphering, Calero2024}.

These various approaches prompt important questions: How effectively do they address AI-related risks, ensure accountability, and foster responsible innovation? Additionally, what criteria should be used for assessing their success, and how can these frameworks adapt for various regional and sectoral contexts? This paper seeks to compare and evaluate AI governance frameworks in the EU, US, UK, and China. This study aims to analyze the strengths and weaknesses of risk mitigation and ethical deployment strategies, establish evaluation criteria for AI risk management, and investigate the growing need for robust AI governance in various global industries and societies. This study uses a multi-method research design, including comparative analysis, thematic analysis, and case study evaluations, along with a thorough literature review to achieve its objectives. This study seeks to enhance current discussions on AI governance and provide concrete insights for developing effective, future-oriented regulatory strategies.

\section{Literature Review}\label{sec:litreview}

\subsection{Evolution of AI Governance Debates}
Early academic focus on AI governance concentrated on broad ethical issues, like the promotion of transparency, fairness, and accountability in algorithmic decision-making \cite{bostrom2014paths, floridi2022unified}. Initial discussions pointed out the risks related to unregulated AI systems, such as potential biases, discriminatory outcomes, and unexpected social impacts \cite{dignum2018ethics}. The rapid commercialization and implementation of AI, especially in sectors such as healthcare, finance, and social services, has revealed the shortcomings of primarily voluntary or principle-based guidelines \cite{eubanks2018automating, 10.1093/polsoc/puae020}. Researchers began pushing for more concrete regulatory measures, stating the importance of balancing innovation incentives with societal protections \cite{bryan2024balancing, Li2024}.

The evolution has been influenced by notable incidents and controversies, including claims of algorithmic discrimination in hiring processes and the inappropriate use of facial recognition technologies in public surveillance \cite{chen2023ethics, TurnerLee2022}. In response, both industry and civil society groups have demanded clearer legal frameworks to clarify liability, protect individuals’ rights, and maintain public trust \cite{DHS2024}. As a result, current discussions on AI governance have transitioned to structured methodologies that focus on high-risk applications and necessitate enhanced oversight and enforcement mechanisms.

\subsection{Risk-Based Approaches to AI Regulation}
An agreement has grown about the importance of risk-based regulatory models that evaluate AI technologies based on their potential harm or societal impact \cite{eu2021proposal}. Rather than applying uniform standards to all AI systems, these models differentiate between low-risk applications, such as basic data analytics, and high-risk or safety-critical systems, including those used in medical diagnostics or autonomous driving \cite{voigt2017eu}. Regulators want to adjust requirements according to risk levels to prevent restricting innovation in low-risk scenarios, while simultaneously ensuring robust protections in situations where AI-driven decisions may impact fundamental rights or public welfare \cite{edwards2021eu}.

Supporters of risk-based approaches argue that these methods offer more defined compliance routes for industries and ensure more consistent enforcement for governmental bodies \cite{OECD2023}. Critics caution that classifying AI systems by risk may be challenging in rapidly changing fields, as the nature and severity of potential harms can evolve over time. A study conducted by the appliedAI Institute for Europe analyzed more than 100 AI systems, revealing that 18\% were categorized as high-risk, 42\% as low-risk, and for 40\%, it was indeterminate whether they belonged to the high-risk category. This ambiguity underscores the challenges in risk classification and indicates that vague classifications may impede investment and innovation. \cite{liebl2023ai}. Risk-based paradigms have emerged as central to numerous contemporary policy proposals, significantly influencing regulatory discussions across various jurisdictions. The OECD AI Principles have been adopted by member countries and various global partners, establishing a basis for international cooperation and interoperability in AI governance \cite{OECD2023}.

\subsection{The European Union’s Pioneering Role}
The EU has led efforts in establishing formal risk-based regulations for AI. The European Commission introduced the AI Act in April 2021, building on the success of the General Data Protection Regulation (GDPR) in establishing global standards for data protection \cite{voigt2017eu} \cite{eu2021proposal}. This proposal, effective August 1, 2024, with full enforcement by August 1, 2027, classifies AI applications into four risk tiers: unacceptable, high, limited, and minimal, imposing stricter requirements on higher-risk categories \cite{schuett2024risk}.

High-risk systems under the AI Act are required to meet obligations related to transparency, data governance, and post-market monitoring. They are subject to conformity assessments and potential oversight by national supervisory authorities \cite{eu2021proposal}. Researchers suggest that the substantial market size of the EU may lead the AI Act to serve as a de facto global standard, which caused multinational companies to match their practices to EU regulations. The phenomenon known as the "Brussels Effect" suggests that internationally operating firms may adopt EU regulations to maintain market access, thus broadening the AI Act's impact beyond Europe. Questions persist regarding the practical enforcement of the Act, especially considering the need for coordination among various regulatory bodies across member states \cite{Demkova2025}.

\subsection{The Decentralized U.S. Model}
The United States uses a decentralized regulatory framework for AI, characterized by a combination of federal and state-level rules and guidelines, in contrast to the EU's top-down approach \cite{cath2018artificial}. Sector-specific agencies, including the Food and Drug Administration (FDA) and the National Highway Traffic Safety Administration (NHTSA), regulate particular AI applications, such as medical devices and autonomous vehicles \cite{bateman2022us}. Furthermore, numerous states have proposed AI-related legislation addressing issues like facial recognition and algorithmic accountability \cite{morley2023operationalising}.

Recent federal initiatives indicate an increasing focus on the need for clearer guidance. The National Institute of Standards and Technology (NIST) published an AI Risk Management Framework in 2023, offering voluntary standards that can help organizations in identifying and mitigating AI risks \cite{ai2023artificial}. The White House has released a “Blueprint for an AI Bill of Rights,” which defines principles like fairness, privacy, and transparency \cite{OSTP2022}. These initiatives show a growing awareness of AI's societal implications; however, the U.S. system continues to be fragmented, with numerous stakeholders expressing concerns regarding deficiencies in legal protections and enforcement \cite{2TurnerLee2022}.

\subsection{The UK’s Sector-Specific Flexibility}
The UK aims to establish itself as a global leader in “pro-innovation”  AI governance, using a flexible, sector-specific strategy that allows regulators to customize regulations for individual industries \cite{UKGovAIRegulation2024,lords2018ai}. Examples include the Financial Conduct Authority guidelines for AI in financial services and the Medicines and Healthcare Products Regulatory Agency's oversight of AI-driven medical devices \cite{singh2024artificial}.

This decentralized model aims to encourage technological experimentation and rapid scaling, while dealing with potential risks through specialized oversight \cite{Hickman2025}. Critics argue that a loose coordination mechanism may end up in inconsistencies and inadequate oversight in high-risk applications. The Ada Lovelace Institute has raised concerns that the current framework may insufficiently address the complexities and risks related to advanced AI systems, which could lead to regulatory gaps \cite{Davies2023}. Current discussions focus on how important it is for the UK to implement more comprehensive legislation versus continuing its sector-by-sector approach, particularly in light of advancing AI technologies and the nation's goal to sustain competitiveness in the global AI landscape. Some experts support a unified regulatory approach to establish consistent standards across sectors, whereas others argue that the current flexible framework promotes adaptability and innovation \cite{HouseOfLordsLibrary2023}. 

\subsection{China’s Centralized, Control-Oriented Strategy}
The governance framework for AI in China is characterized by state-led directives that integrate AI development with general national objectives in technology, security, and economic growth \cite{Wang2024}. The government has implemented specific regulations for technologies including facial recognition and generative AI, often requiring registration and algorithmic audits \cite{cheng2023shaping}. The Personal Information Protection Law (PIPL) and the Data Security Law regulate data management and global data transfers \cite{chen2024developing}.

This centralized approach may enable fast execution of extensive AI initiatives; however, scholars raise concerns about privacy, civil liberties, and the possibility of exporting of this model to other jurisdictions \cite{DHS2024, creemers2021china}. Recent regulations in China include regulations for real-time monitoring of AI-generated content, aimed at making social stability and national security, in addition to data protection measures. The "Interim Measures for the Management of Generative Artificial Intelligence Services," which took effect in August 2023, require that AI-generated content follow with the Core Socialist Values and avoid producing material that may disrupt economic or social order. Providers have to use real-time monitoring tools and data analysis techniques to detect and address abnormal activities or content generated by AI systems \cite{ye2024privacy}.

\subsection{Gaps in Comparative Analyses}
Despite a large amount of research on individual AI governance frameworks, there is a notable lack of systematic cross-regional comparisons of AI risk management frameworks \cite{luna2024navigating}. Research frequently confines itself to descriptive analyses of legislation or general ethical principles, ignoring to investigate the practical implementation of these policies or to offer sector-specific comparisons \cite{schmitz2024global}. The literature rarely offers consistent criteria—such as risk mitigation, adaptability, transparency, and implementation feasibility—for assessing various regulatory approaches \cite{batool2025ai}. The growing number of AI applications across different industries requires integrated analyses to inform policymakers, industry stakeholders, and civil society.

\subsection{Toward a Comprehensive Comparative Framework}
This paper aims to synthesize existing scholarship and regulatory documents to offer a comparative study of AI governance in the EU, U.S., UK, and China. The research uses qualitative methods, including thematic analysis, case studies, and framework-based evaluation, to explain how different governance models respond to the risks and opportunities presented by AI. This study aims to enhance the discussion on effective and context-sensitive AI regulation by addressing both high-level legal frameworks and practical implementation challenges. The findings aim to inform future policy decisions, encourage international collaboration, and make sure the responsible use of AI technologies for collective benefit.

\section{Methodology}\label{sec:methodology}

This research uses a multi-method qualitative approach to systematically analyze AI risk management frameworks in the four regions. The research integrates comparative policy analysis, thematic analysis, and case study methodology to develop a nuanced understanding of how these frameworks address different aspects of AI risk and governance.

\subsection{Research Design}
The research design has been designed to take into account both the extensive and intensive aspects of AI governance approaches.

\begin{figure}[htbp!]
    \centering
    \resizebox{0.5\textwidth}{!}{\includegraphics{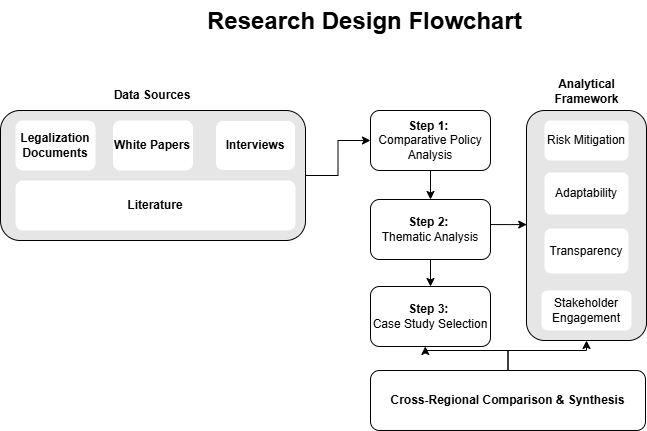}} % Adjust 0.8 for size
    \caption{Research Design Flowchart}
    \label{fig:research_design}
\end{figure}

\begin{enumerate}
    \item \textbf{Comparative Policy Analysis:} This paper examines legislative texts, regulatory guidelines, and enforcement mechanisms across four regions, referencing established cross-jurisdictional studies \cite{hantrais2008international}. This comparison discusses the similarities and differences in the classification, monitoring, and enforcement of risks.
    \item \textbf{Thematic Analysis:} Important themes, including accountability, transparency, adaptability, and stakeholder engagement, were identified through an initial review of academic and policy literature \cite{braun2006using}. Themes informed the coding of policy documents, uncovering underlying assumptions and governance priorities within each jurisdiction.
    \item \textbf{Case Study Selection:} The study analyzes specific high-risk domains across various regions to understand real-world applications. Domains were selected due to their potential societal impact and the presence of publicly documented policy interventions \cite{stake1995case}. Case examples provided in a later section demonstrate the practical functioning of regulations, including enforcement challenges.
\end{enumerate}
\autoref{fig:research_design} shows a visual overview of the research design.

\subsection{Data Sources}
Data collection focused on four categories. An overview is provided in \autoref{tab:datasources}.

\begin{table*}[htbp] % Dynamically spans columns
\centering
\caption{Overview of Data Sources}
\label{tab:datasources}
\resizebox{\textwidth}{!}{%
\begin{tabular}{p{4.5cm} p{6cm} p{6cm}}
\toprule
\textbf{Category} & \textbf{Examples} & \textbf{Rationale / References} \\
\midrule
\textbf{Legislation and Policy Documents} & AI Act (EU), State and Federal Regulations (U.S.), Sector-Specific Guidance (UK), Government Directives (China) & Primary source for official rules and obligations \cite{eu2021proposal} \\
\textbf{Academic Literature} & Peer-reviewed journals, conference proceedings & Provides theoretical context and empirical insights \cite{morley2023operationalising} \\
\textbf{Industry White Papers} & Company guidelines, professional association reports & Reflects industry perspectives and compliance strategies \cite{ai2023artificial} \\
\textbf{Interviews} & Legal experts, policymakers, industry stakeholders & Offers qualitative insights into practical implementation and ongoing policy debates \cite{creswell2016qualitative} \\
\bottomrule
\end{tabular}
}
\end{table*}

\subsection{Analytical Framework}
Building on prior research into AI governance and policy evaluation, this study uses the criteria shown in \autoref{tab:criteria} to assess each regulatory framework. Each dimension is grounded in the literature and responds to recognized gaps in existing comparative analyses \cite{wu2023comprehensive}.

\begin{table*}[htbp] % Dynamically spans columns
\centering
\caption{Criteria for Evaluating AI Governance Frameworks}
\label{tab:criteria}
\resizebox{\textwidth}{!}{%
\begin{tabular}{p{4.5cm} p{12cm}}
\toprule
\textbf{Criterion} & \textbf{Description} \\
\midrule
\textbf{Risk Mitigation Capability} & Assesses whether frameworks effectively address ethical, security, and societal harms (e.g., bias, privacy breaches). \\
\textbf{Regulatory Comprehensiveness} & Examines the breadth of AI applications covered and the specificity of implementation guidelines (e.g., high-risk focus, enforcement mechanisms). \\
\textbf{Adaptability} & Evaluates the capacity of regulations to evolve in response to emerging AI technologies and changing risk profiles. \\
\textbf{Transparency \& Accountability} & Investigates requirements for explainability, auditability, and oversight in high-risk AI systems (e.g., mandated disclosures, independent audits). \\
\textbf{Stakeholder Engagement} & Considers the involvement of civil society, industry, and academia in shaping and refining the regulatory framework (e.g., consultation, public comment periods). \\
\bottomrule
\end{tabular}
}
\end{table*}

\noindent
The analysis sections evaluate each jurisdiction's approach against mentioned criteria, allowing a systematic cross-regional comparison that highlights both convergent and divergent policy strategies.

\subsection{Scope \& Limitations}
This study focuses on formal governance instruments related to high-risk AI applications, where public safety and individual rights are significantly at risk. The multi-method approach provides both breadth and depth; however, it is important to recognize several limitations. Regulatory flux presents a challenge, as AI regulations are evolving rapidly. Ongoing legislative updates, including new rules for generative AI, could make parts of this analysis outdated. Language and translation issues come out when analyzing policy documents from non-English-speaking contexts, especially in China, where translations may exclude nuanced legal or cultural interpretations \cite{lost_in_translation}. The contextual complexity of AI governance is significant as cultural, political, and economic factors significantly impact regulatory models, therefore limiting direct comparisons across jurisdictions. Data availability continues to pose a constraint, as the selection of case studies and interviews relies on part on publicly accessible sources, potentially introducing bias if certain industry or policy perspectives are underrepresented.

Despite these constraints, the selected methodology facilitates a context-aware review of AI risk management practices, with the objective of contributing to both academic discussions and practical policy considerations about the regulation of rapidly evolving AI technologies.

\section{Comparative Analysis and Evaluation of Effectiveness}

\subsection{Risk Categorization \& Mitigation}\label{sec:riskcat_mitigation}

A significant challenge in AI governance includes classifying various AI applications based on their potential ethical, security, and societal risks, followed by the development of measures to mitigate these risks \cite{OECD2023}. The methodology outlined in \autoref{sec:methodology} presents an analytical framework that evaluates the regulatory scope and enforcement strategies of each jurisdiction. This section analyzes the risk categorization frameworks used by the regions, and evaluates their effectiveness in addressing identified risks.

\paragraph{European Union (EU)}
The EU's proposed AI Act presents a structured framework for risk categorization, defining four distinct tiers: unacceptable, high, limited, and minimal \cite{eu2021proposal}. High-risk systems, including those used in healthcare diagnostics, biometric identification, and critical infrastructure, are required to adhere to regulations related to data governance, transparency, and post-market monitoring \cite{EUAIArticle72,Lumenova2024}. 

Ethical risks come out from the Act's requirement for developers to document algorithmic decisions and perform ex-ante assessments to identify biases or discriminatory outcomes. This regulation encourages accountability and facilitates the potential for external audits \cite{voigt2017eu}.

Security risks are addressed via conformity assessments and cybersecurity standards, developed to ensure that high-risk systems are strong to attacks or manipulation. Providers are required to maintain technical documentation and logs to facilitate incident investigation \cite{schuett2024risk}.

Transparency obligations, such as notifying users during their interactions with AI, serve to mitigate societal risks. The measures aim to enhance public trust and mitigate societal harms, such as disinformation and violations of fundamental rights \cite{schuett2024risk}.

The tiered obligations clarify compliance pathways; however, effective mitigation relies significantly on consistent enforcement and collaboration among member states. Due to the differing capabilities of national regulators, there are ongoing concerns regarding the consistent implementation across the EU \cite{Zaidan2024}.

\paragraph{United States (U.S.)}
The U.S. uses a decentralized a, with federal agencies identifying AI-related risks specific to their domains instead of relying on a unified classification framework \cite{cath2018artificial}. The Food and Drug Administration (FDA) regulates AI-driven medical devices through classification based on patient safety implications, whereas the National Highway Traffic Safety Administration (NHTSA) points out the safety of autonomous vehicles \cite{morley2023operationalising}. 

Self-regulation within the industry frequently dominates the discussion on ethical risks in AI governance. Although certain federal guidelines offer recommendations about fairness and nondiscrimination \cite{OSTP2022}, they do not include binding enforcement mechanisms, leading to compliance being primarily voluntary.

Security risks associated with AI applications can be reduced by implementing specific cybersecurity requirements within critical infrastructure sectors. However, a unified standard applicable to all high-risk AI applications is lacking \cite{Hickman2025}. Recent initiatives, among them the NIST AI Risk Management Framework, advocate for best practices in AI security; however, their implementation is still voluntary \cite{ai2023artificial}.

Societal risks arise from the lack of a comprehensive national strategy for algorithmic accountability, which leads to uneven regulation of AI technologies. Despite attempts to establish a cohesive federal framework, no legislation has been passed. The Algorithmic Accountability Act of 2023 aims to enable the Federal Trade Commission to require impact assessments for automated decision systems and significant decision-making processes \cite{AlgorithmicAccountabilityAct2023}, particularly concerning facial recognition and social media content moderation \cite{Lively2021} . Although these measures look for to encourage innovation, societal harms may continue unless mitigated by more localized or sector-specific regulations.

The U.S. framework demonstrates a capacity for rapid adaptation to emerging technologies through specialized agencies; however, fragmented governance may result in notable coverage gaps, especially regarding ethical and societal risks.

\paragraph{United Kingdom (UK)}
The strategy used by the UK relies on sector-specific guidance instead of a unified legislative framework for the classification of AI risks \cite{singh2024artificial}. Regulatory bodies such as the Financial Conduct Authority (FCA) and the Medicines and Healthcare products Regulatory Agency (MHRA) establish domain-specific requirements, using a proportional approach that adjusts risk thresholds based on sector variations \cite{lords2018ai}.

Ethical risks emerge when regulators establish guidelines or codes of practice that may include metrics for fairness and accountability for AI developers. The lack of a universal framework may result in inconsistencies in the implementation of ethical considerations across various sectors.

Security risks are mainly addressed through established data protection and cybersecurity regulations, such as the Data Protection Act \cite{DataProtectionAct2018}, which are implemented in a case-by-case manner for AI applications. This approach permits flexibility but may lead to ambiguity concerning the ultimate responsibility for managing cross-sectoral AI threats \cite{Hickman2025}.

Societal risks are managed through public consultations and expert committees, including the Centre for Data Ethics and Innovation, which offer insights on wider societal issues, such as labor displacement and algorithmic discrimination \cite{lords2018ai}. In the absence of comprehensive legislation, the implementation of these recommendations varies significantly among industries.

Supporters argue that the UK's framework encourages adaptive governance and quick sectoral revisions, whereas critics emphasize the risks of regulatory fragmentation and inadequate protections against systemic AI threats \cite{Hickman2025}.

\paragraph{China}
In China, risk categorization frequently corresponds with governmental priorities regarding social stability and economic development \cite{ding2018deciphering}. Government entities regularly release directives that specify "key areas" (e.g., facial recognition, online content moderation) in which AI developers are required to adhere to stricter regulations \cite{ChinaSocialCredit}.

Official documents increasingly recognize ethical risks within the framework of "ethical AI." Implementation efforts primarily concentrate on ensuring that content and algorithmic outputs follow social and political norms \cite{LathamWatkins2023}. Current mechanisms prioritize internal audits and adherence to "core socialist values" over the establishment of independent oversight.

Mandatory registration and algorithmic audits are implemented to mitigate security risks, with an eye on the prevention of data leaks and malicious manipulation. Cybersecurity legislation includes AI applications, especially in scenarios including national security or public safety \cite{creemers2021china}.

AI-based surveillance presents societal risks that are regulated by comprehensive data governance laws, such as the Data Security Law (DSL), aimed at preserving public order \cite{ForeignPolicy2022}. However, opportunities for citizens and civil society actors to contest or appeal decisions related to AI are currently restricted.

China's centralized and control-oriented approach allows rapid policy implementation; however, international commentators have expressed concerns about transparency, potential overreach, and its effect on individual rights \cite{DHS2024}.

\paragraph{Comparative Assessment of Risk Mitigation Effectiveness}
A cross-jurisdictional comparison (see \autoref{tab:riskcompare}) reveals distinct advantages and shortcomings in each framework’s capacity to mitigate ethical, security, and societal risks.

\begin{table*}[htbp] % Dynamically spans columns
\centering
\caption{Comparative Overview of Risk Categorization \& Mitigation}
\label{tab:riskcompare}
\resizebox{\textwidth}{!}{%
\begin{tabular}{p{3.2cm} p{3.2cm} p{3.2cm} p{3.2cm} p{3.2cm}}
\toprule
\textbf{Dimension} & \textbf{EU} & \textbf{U.S.} & \textbf{UK} & \textbf{China} \\
\midrule
\textbf{Risk Model} & Tiered (4 levels) & Decentralized, sector-driven & Sector-specific, flexible & Centralized directives aligned with state priorities \\
\textbf{Ethical Risks} & Mandatory bias checks, transparency & Mostly voluntary; agency-specific & Guidance-based; uneven adoption & Internal audits \& alignment with state values \\
\textbf{Security Risks} & Conformity assessments, logs & Varied agency standards, no uniform approach & Data protection laws apply case-by-case & Mandatory registration, cybersecurity focus \\
\textbf{Societal Risks} & Transparency obligations, user rights & Limited federal oversight; patchwork rules & Public consultation \& ethical committees & State oversight to maintain social stability \\
\textbf{Key Strength} & Clear legal framework, strong procedural safeguards & Flexibility \& sectoral expertise & Agility \& industry-specific adaptation & Rapid implementation \& enforcement \\
\textbf{Primary Gap} & Enforcement uniformity across member states & Fragmentation \& uneven protection & Risk of regulatory fragmentation & Potential overreach; limited public recourse \\
\bottomrule
\end{tabular}
}
\end{table*}

\FloatBarrier % Ensures text doesn’t interfere with the table

\begin{figure}[htbp]
    \centering
    \resizebox{0.4\textwidth}{!}{\includegraphics{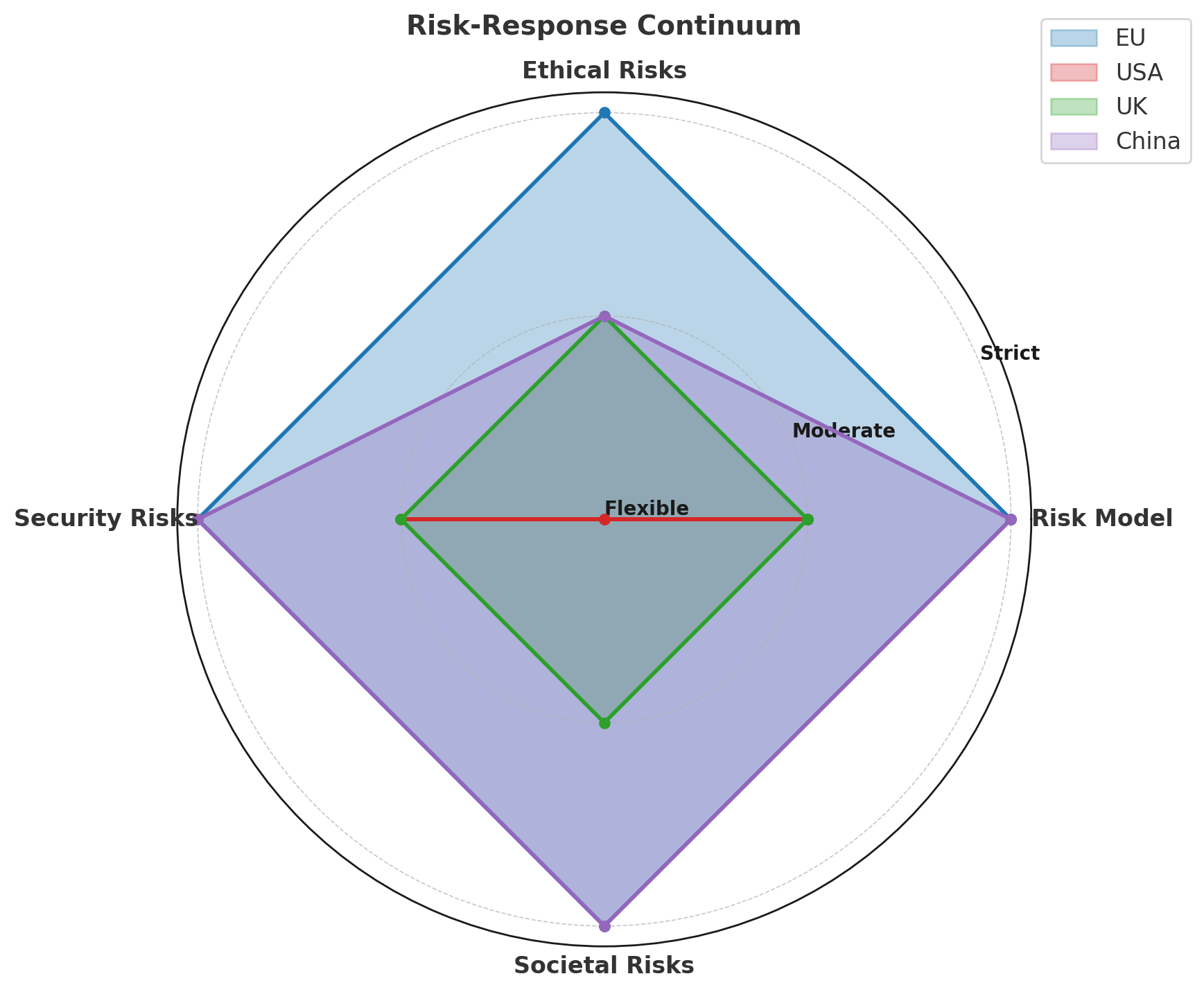}} % Adjust 0.8 for size
    \caption{AI Risk Management Strictness Across Jurisdictions}
    \label{fig:risk_response_continuum}
\end{figure}

The EU stands out for its structured and legally binding definitions, which determine a solid foundation for managing high-risk AI. Practical enforcement will depend on the coordination of oversight resources among member states. The U.S. model encourages quick innovation while revealing deficiencies in regulatory oversight, particularly concerning ethical and societal implications. The UK's ability to keep up with sector-specific guidance is beneficial; however, it may result in inconsistencies among industries. China's centralized framework effectively addresses security risks but may undermine individual autonomy and oversight mechanisms that are important in more liberal democracies. \autoref{fig:risk_response_continuum} illustrates the variation in regulatory responses across jurisdictions by highlighting the relative strictness of each region's approach for different AI risk levels.

Each jurisdiction balances innovation with risk mitigation differently. While all recognize the need to classify AI systems by risk level, they diverge in how they define, operationalize, and enforce those classifications. These findings align with previous literature suggesting that cultural, political, and economic factors profoundly shape AI governance strategies \cite{creemers2021china}.

\subsection{Governance \& Oversight}\label{sec:gov_oversight}

Regulatory frameworks for AI risk management establish rules and technical requirements, while also defining the entities responsible for compliance oversight and the methods of using such oversight. Effective governance generally covers a collaboration among government agencies, industry participants, and civil society organizations, with each playing specific roles in policy development, oversight, and enforcement \cite{morley2023operationalising, cath2018artificial}. This criteria compares the governance models across the four regions, analyzing their oversight mechanisms and evaluating their practical effectiveness.

\paragraph{\textbf{European Union (EU)}}
The EU centralizes regulatory policymaking through institutions like the European Commission and the European Parliament, which work together on legislative initiatives, including the AI Act \cite{eu2021proposal}. Upon implementation, national competent authorities in each member state will conduct market surveillance, investigate instances of noncompliance, and impose sanctions as needed \cite{Zenner2024}. Industry stakeholders, especially large technology firms, frequently participate in consultations or pilot projects to offer insights on regulatory feasibility and potential economic impacts \cite{TimeThierryBreton2024}. Civil society organizations (CSOs) and think tanks serve an essential advisory work, encouraging ethical AI principles and highlighting social risks \cite{EPLegislativeTrain2024}.

The multi-tiered governance structure of the EU facilitates comprehensive oversight; however, coordination between EU institutions and national authorities may result in enforcement inconsistencies \cite{smith2018challenges}. Although established frameworks for stakeholder engagement are available, smaller businesses and underfunded NGOs often struggle to compete with large corporations in influencing policy discussions \cite{Demirci2024}.

\paragraph{United States (U.S.)}
In the U.S., AI governance is implemented through a combination of federal agencies, such as the Food and Drug Administration, Federal Trade Commission, and National Highway Traffic Safety Administration, alongside state-level organizations \cite{ai2023artificial}. Self-regulation within the industry is common, as major technology companies establish internal ethics boards or publish ethical guidelines \cite{OSTP2022}. Non-governmental organizations (NGOs) and advocacy groups often engage in lobbying efforts aimed at enhancing consumer protections and transparency measures, with their influence fluctuating based on political and economic contexts \cite{OpenSecrets2024}.

This decentralized model allows specialized agencies to utilize domain-specific expertise, potentially facilitating more agile responses to emerging AI applications \cite{morley2020initial}. Overlaps or gaps between federal and state regulators can create ambiguity regarding the entity with ultimate oversight authority. The role of civil society, although strong in certain policy discussions, is occasionally eclipsed by well-funded technology companies that can greatly influence regulatory priorities \cite{cath2018artificial}.

\paragraph{United Kingdom (UK)}
In the UK, governance of AI is characterized by a sector-specific approach, with regulatory bodies like the Financial Conduct Authority (FCA), the Information Commissioner’s Office (ICO), and the Medicines and Healthcare products Regulatory Agency (MHRA) responsible for overseeing AI applications within their respective fields \cite{WhittakerSmith2024}. The Centre for Data Ethics and Innovation (CDEI) provides guidance to the government on ethical and societal matters, promoting public consultations and evidence-based policymaking \cite{CDEI2025}.

The sectoral approach uses specialized regulatory knowledge, allowing adaptable responses to emerging AI technologies \cite{Hickman2025}. However, the absence of comprehensive legislation similar to the EU’s AI Act could result in differences in oversight among various industries. Civil society frequently engages in policymaking through consultations; however, industry has significant influence, particularly in emerging areas that lack established guidelines \cite{Kaleka2024}.

\paragraph{China}
China's governance of AI is defined by a state-centric approach, with major organizations like the Cyberspace Administration of China and the Ministry of Industry and Information Technology leading policy directives and enforcement \cite{ding2018deciphering}. Major technology firms frequently engage in close collaboration with government agencies, aligning their business operations with national strategies for the advancement of artificial intelligence and social governance \cite{LOC2023}. Civil society participation is relatively restricted, with state-sanctioned organizations assuming a more significant role than independent NGOs \cite{ICNL2024}.

Centralized oversight allows the government to rapidly implement and enforce regulations, particularly in critical domains such as surveillance and public services \cite{creemers2021china}. Critics argue that a top-down system offers limited opportunities for independent audits or public accountability, which may compromise personal liberties and restrict external scrutiny of AI implementations \cite{ding2018deciphering}.

\paragraph{Comparative Assessment of Governance Structures}
\autoref{tab:governance_comparison} summarizes the roles of government, industry, and civil society across the four jurisdictions, along with the effectiveness of each structure in practice.

\begin{table*}[htbp] % Dynamically spans columns
\centering
\caption{Governance \& Oversight Structures Across Jurisdictions}
\label{tab:governance_comparison}
\resizebox{\textwidth}{!}{%
\begin{tabular}{p{4cm} p{4.5cm} p{4.5cm} p{4.5cm} p{4.5cm}}
\toprule
\textbf{Dimension} & \textbf{EU} & \textbf{U.S.} & \textbf{UK} & \textbf{China} \\
\midrule
\textbf{Government Role} & \parbox{4.5cm}{\raggedright Central EU institutions + Member State authorities} & \parbox{4.5cm}{\raggedright Federal agencies + State-level bodies} & \parbox{4.5cm}{\raggedright Sector-specific regulators} & \parbox{4.5cm}{\raggedright Centralized, top-down oversight} \\
\textbf{Industry Role} & \parbox{4.5cm}{\raggedright Consultation, compliance} & \parbox{4.5cm}{\raggedright Self-regulation, lobbying} & \parbox{4.5cm}{\raggedright Guidance collaboration} & \parbox{4.5cm}{\raggedright Close alignment with state policy} \\
\textbf{Civil Society Role} & \parbox{4.5cm}{\raggedright Advisory, watchdog activities} & \parbox{4.5cm}{\raggedright Advocacy, limited success in certain sectors} & \parbox{4.5cm}{\raggedright Advisory committees, public consultations} & \parbox{4.5cm}{\raggedright Restricted, often state-sanctioned} \\
\textbf{Effectiveness} & \parbox{4.5cm}{\raggedright Potential for comprehensive coverage but uneven enforcement} & \parbox{4.5cm}{\raggedright Deep expertise but fragmented oversight} & \parbox{4.5cm}{\raggedright Flexible adaptation, risk of inconsistency} & \parbox{4.5cm}{\raggedright Rapid implementation, possible overreach} \\
\bottomrule
\end{tabular}
}
\end{table*}

Overall, each jurisdiction orchestrates a distinct balance among government bodies, industry actors, and civil society. While the EU’s structured approach and China’s centralized model offer strong top-down regulation, both raise concerns about consistent or transparent enforcement. The U.S. and UK rely more on decentralized oversight, using specialized agencies and market-driven solutions, but risk patchy or inconsistent coverage. Civil society’s engagement varies widely, shaped by historical, political, and cultural contexts in each region.

\subsection{Transparency \& Explainability}\label{sec:transparency_explainability}

A fundamental principle of responsible AI governance is the need for transparency and explainability in both technical processes and decision-making outcomes. Transparency is related to the clarity about the functionality and data used by AI systems, while explainability signifies the ability of stakeholders—such as regulators, end-users, or affected individuals—to comprehend the processes underlying AI-generated decisions \cite{F52024}. This section analyzes the approaches of the European Union (EU), United States (U.S.), United Kingdom (UK), and China regarding transparency and explainability in their regulatory or policy frameworks, and assesses the practical implications for stakeholder trust.

\paragraph{European Union (EU)}
The EU’s proposed AI Act requires that developers of high-risk AI systems provide comprehensive documentation and technical specifications to demonstrate adherence to applicable standards, along with user-facing disclosures during any interaction with AI \cite{eu2021proposal}. The requirements extend previous standards established by the General Data Protection Regulation (GDPR), which includes provisions for data subjects to access and amend automated decisions \cite{gdpr2022automated}. The AI Act requires conformity assessments that emphasize transparency and auditability, thereby improving ex ante accountability mechanisms. 

These measures enhance trust by providing consumers with greater insight into AI processes; however, some critics argue that the documentation may be overly technical or fragmented for non-expert stakeholders \cite{morley2020initial}. Small and medium-sized enterprises (SMEs) encounter difficulties in adhering to complex explainability standards, which may exacerbate the disparity between larger and smaller market participants \cite{Schwaeke13082024, schmude2025information}.

\paragraph{United States (U.S.)}
The U.S. does not have comprehensive legal requirements for algorithmic transparency or explainability beyond specific sectors. The Food and Drug Administration (FDA) occasionally requires that manufacturers of AI-powered medical devices provide data elucidating diagnostic algorithms \cite{FDA2025}. In consumer-facing applications, numerous companies voluntarily publish ethics guidelines or transparency reports; however, these practices largely lack regulation \cite{OSTP2022}. The Federal Trade Commission (FTC) has indicated that unfair or deceptive practices concerning opaque AI might violate consumer protection laws, despite the absence of explicit regulations at this time \cite{FTC2025}.

The focus on industry self-regulation could encourage innovation in explainable AI (XAI) techniques; however, it also creates vulnerabilities that can undermine stakeholder trust, especially in domains like facial recognition and credit scoring, where users possess limited means to contest AI-generated results \cite{Jones2021, Guardian2024}.

\paragraph{United Kingdom (UK)}
The UK allocates transparency and explainability requirements across multiple regulatory bodies, mirroring its sector-specific approach. The Information Commissioner’s Office (ICO) establishes guidelines for data protection and automated decision-making, encouraging organizations to implement “meaningful information” disclosure \cite{ICO2025}. In regulated sectors such as finance and healthcare, sectoral regulators offer supplementary guidance on the interpretation of transparency obligations within their respective domains \cite{GLI_BankingUK2024}.

The UK model includes transparency requirements within existing regulatory frameworks, allowing contextual adaptation. However, the absence of a cohesive AI-specific legal framework, exemplified by the EU’s AI Act, results in difficulties for certain stakeholders in distinguishing between legally required disclosures and those that are just recommended \cite{roberts2023artificial}. This may erode trust if users see transparency measures as inconsistent or inadequately enforced.

\paragraph{China}
China's regulatory documents increasingly recognize the importance of algorithmic transparency; however, policies frequently prioritize alignment with national security and "core socialist values" over public-facing disclosure \cite{ChinaAIInterim2023}. Regulations from the Cyberspace Administration of China (CAC) and other governmental entities typically mandate that high-risk AI systems, particularly those utilized for content generation or social media, undergo internal audits. Developers may be required to provide comprehensive information about their algorithms and data sources to regulatory bodies; however, there is insufficient focus on elucidating automated decisions to end-users \cite{ChinaAIInterim2023, menglu2024regulation}.

This internal and state-led model effectively addresses concerns related to "undesirable" or destabilizing content; however, it provides limited transparency to the public concerning the decision-making and moderation processes. Therefore, stakeholder trust is significantly dependent on governmental authority, with limited independent avenues for validating or critiquing AI-driven results \cite{ChinaAIInterim2023}.

\paragraph{Comparative Assessment of Transparency \& Explainability}
\autoref{tab:transparency_explain} illustrates how different regulatory approaches promote or neglect transparency and explainability, ultimately shaping stakeholder trust.

\begin{table*}[htbp] % Dynamically spans columns
\centering
\caption{Comparison of Transparency \& Explainability Initiatives}
\label{tab:transparency_explain}
\resizebox{\textwidth}{!}{%
\begin{tabular}{p{3.5cm} p{4.5cm} p{4.5cm} p{4.5cm}}
\toprule
\textbf{Jurisdiction} & \textbf{Key Transparency Mechanisms} & \textbf{Extent of Explainability} & \textbf{Impact on Stakeholder Trust} \\
\midrule
\textbf{EU} & Mandatory user disclosures, conformity assessments & Strong formal requirements, though potentially technical & Generally high trust, but SMEs may struggle with compliance \\
\textbf{U.S.} & Mostly voluntary; sector-specific rules (FDA, FTC) & Uneven; dependent on self-regulation & Encourages innovation but risks trust deficits in opaque applications \\
\textbf{UK} & Distributed via ICO \& sector regulators & “Contextual” explanation; no single statutory standard & Variable trust; clarity depends on industry guidelines \\
\textbf{China} & Internal audits submitted to authorities & Public-facing explanations often limited & Trust aligns with state oversight, fewer independent validation channels \\
\bottomrule
\end{tabular}
}
\end{table*}

In particular, the EU's comprehensive approach -exemplified by the AI Act and its conformity assessment procedures - positions transparency as a legal obligation, potentially strengthening trust but imposing significant compliance demands. In the U.S. and UK, market-driven or sector-driven approaches can foster tailored transparency measures yet risk inconsistent protection for end-users. China’s system emphasizes internal reporting and aligns explainability with national policy objectives, offering limited visibility to independent actors.

In general, transparency and explainability remain essential for cultivating stakeholder trust across jurisdictions. Yet each region operationalizes these principles differently, reflecting broader legal and cultural frameworks. Future developments may involve continued experimentation with explainable AI (XAI) techniques, policy refinements to address technical complexity, and greater collaboration between governments, industry, and civil society to make transparency and explainability more accessible and actionable for diverse stakeholders.

\subsection{Adaptability \& Innovation}\label{sec:adapt_innovation}

As advancements in generative AI, autonomous vehicles, and advanced robotics speed up, the capacity of regulatory frameworks to adapt to these emerging technologies is an important part of their effectiveness \cite{llorca2024testing, DigitalRegulation_AI_Challenges_2024}. This section evaluates the approaches of the European Union (EU), United States (U.S.), United Kingdom (UK), and China in adapting to technological evolution and analyzes their effectiveness in harmonizing innovation with stringent regulation.

\paragraph{European Union (EU)}
The EU's AI Act includes elements of adaptability, allowing for updates to the classification of high-risk applications through delegated acts \cite{eu2021proposal}. This mechanism allows policymakers to incorporate newly identified risk areas without the need for a completely new regulation. The Act emphasizes risk-based proportionality, offering flexibility in accommodating different AI systems.

The EU model enhances legal certainty and establishes a global benchmark; however, businesses, particularly start-ups, are apprehensive that compliance costs might prevent experimentation and innovation \cite{ITIF2021}.  
The European Innovation Council and Horizon Europe programs aim to help solve these challenges through funding for AI research and development; however, tensions remain between the desire to lead in AI regulation and the potential to discourage risk-taking initiatives \cite{Orrick2024}.

\paragraph{United States (U.S.)}
The U.S. approach, characterized by fragmentation and sector-specific focus, allows for rapid adaptation to emerging AI applications within particular industries. Federal and state regulators frequently issue or revise guidelines in reaction to technological advancements; for instance, the National Institute of Standards and Technology (NIST) regularly updates its voluntary frameworks to incorporate emerging best practices \cite{ai2023artificial}.

The U.S. model encourages rapid implementation and market-driven experimentation by avoiding a singular comprehensive statute \cite{ai2023artificial}. Technology firms identify this environment as being helpful to promoting global leadership in AI.  
The lack of a cohesive national framework may lead to regulatory uncertainty for developers working in various states or sectors. This additionally raises concerns about the insufficient regulation of novel technologies that do not align with current agency mandates \cite{cath2018artificial, morley2020initial}.

\paragraph{United Kingdom (UK)}
The UK's sector-specific and flexible approach allows regulators to quickly integrate new guidelines as AI develops \cite{lords2018ai}. Organizations such as the Centre for Data Ethics and Innovation (CDEI) systematically observe technological developments and are capable of recommending specific modifications to industry regulations \cite{UKGovAIRegulation2024}.

Supporters contend that this agility makes the UK to respond quickly to emerging challenges, like general AI models \cite{Bhatti2024}.  
Critics warn that the absence of comprehensive legislation may lead to inconsistent oversight, potentially causing significant issues when a disruptive AI application crosses various regulatory domains \cite{Kaleka2024}.

\paragraph{China}
The regulatory model in China, characterized by centralization and directive-based governance, allows for the quick revision or issuance of regulations in alignment with evolving strategic or security priorities \cite{ding2018deciphering}. Authorities quickly implemented regulations for generative AI, requiring real-time monitoring and compliance with state-designated values \cite{ChinaAIInterim2023}.

Public-private collaborations with a large scale, supported by considerable government funding, have driven China's growth in AI research and commercialization \cite{Bhatti2024}.  
Critics argue that strict oversight and a focus on national security can limit open-ended innovation, especially in domains that could challenge political or social norms \cite{zhang2024promise}.

\paragraph{Balancing Innovation \& Robust Regulation}
\autoref{tab:adapt_innovation} shows how each jurisdiction handles adaptability and the trade-off between fostering innovation and safeguarding societal interests.

\begin{table*}[htbp]
\centering
\caption{Comparison of Adaptability \& Innovation Across Jurisdictions}
\label{tab:adapt_innovation}
\resizebox{\textwidth}{!}{%
\begin{tabular}{p{3cm} p{5cm} p{5cm} p{5cm}}
\toprule
\textbf{Jurisdiction} & \textbf{Adaptability Features} & \textbf{Innovation Incentives} & \textbf{Potential Drawbacks} \\
\midrule
\textbf{EU} & \parbox{5cm}{\raggedright Delegated acts to update AI Act \\ Risk-based proportionality} & 
\parbox{5cm}{\raggedright Research grants (Horizon Europe) \\ Structured compliance} & 
\parbox{5cm}{\raggedright Higher entry barrier for SMEs \\ Possible overregulation} \\
\textbf{U.S.} & \parbox{5cm}{\raggedright Agency rulemaking \\ State-level experimentation} & 
\parbox{5cm}{\raggedright Market-driven \\ Minimal ex-ante constraints} & 
\parbox{5cm}{\raggedright Fragmented coverage \\ Regulatory gaps for emerging tech} \\
\textbf{UK} & \parbox{5cm}{\raggedright Sector-specific guidelines \\ Expert committees (CDEI)} & 
\parbox{5cm}{\raggedright Light-touch approach \\ Rapid guidance updates} & 
\parbox{5cm}{\raggedright Inconsistent standards \\ Potential confusion across sectors} \\
\textbf{China} & \parbox{5cm}{\raggedright Centralized directives revised swiftly} & 
\parbox{5cm}{\raggedright Significant public investment \\ Close industry-government coordination} & 
\parbox{5cm}{\raggedright Strict oversight may limit open innovation \\ Focus on national priorities} \\
\bottomrule
\end{tabular}
}
\end{table*}

Adaptability and innovation are crucial parts of governance. The EU's structured and uniform model offers predictability; however, it may hinder agile experimentation. The US prioritizes flexibility and market-driven growth; however, this approach may lead to oversight gaps and legal uncertainties. The sector-specific system in the UK provides a compromise, though it may lead to inconsistencies and coordination difficulties. The rapid, state-led regulatory changes in China can expedite the development of emerging technologies in prioritized sectors, while simultaneously diminishing openness and pluralism within the innovation ecosystem.

From a policy standpoint, achieving an appropriate balance between encouraging AI innovation and maintaining effective oversight continues to be an ongoing obstacle across all four jurisdictions. Policymakers are refining regulations to address emerging technological frontiers, including generative AI, cognitive robotics, and quantum-based machine learning, underscoring the evolving nature of AI governance in various global contexts.

\section{Case Studies}\label{sec:casestudies}

To illustrate how AI governance frameworks function in concrete settings, this section presents concise case studies from each of the four jurisdictions examined. These real-world examples show the interaction between regulatory requirements, practical challenges, and effectiveness in mitigating risks \cite{stake1995case}.

\subsection{European Union: Healthcare Diagnostics}
A major University hospital in Germany adopted an AI-driven diagnostic tool for radiology, classifying chest X-rays to detect early signs of pneumonia and other lung diseases \cite{deepc2023}. Given the tool’s high-risk healthcare application, it fell under the EU AI Act’s stricter compliance requirements \cite{eu2021proposal}.

Bias and data quality posed significant concerns, as the AI model risked misdiagnosing underrepresented patient groups if the training data lacked sufficient diversity \cite{James2024}.

Transparency and post-market monitoring requirements mandated that the hospital maintain technical documentation and demonstrate continuous monitoring in compliance with the AI Act \cite{healthCareServices}.

Patient privacy was another critical issue, as the General Data Protection Regulation (GDPR) provisions required the implementation of robust data protection measures, adding an additional layer of compliance \cite{sartor2020impact}.

The integration of an AI-driven diagnostic tool within a German hospital network has yielded notable improvements in diagnostic accuracy, particularly in the early detection of common conditions. This advancement has the potential to alleviate healthcare professionals' workloads and reduce overall operational costs. For instance, AI applications in radiology have demonstrated a reduction in diagnostic time by approximately 90\%, significantly enhancing efficiency \cite{Jeong2025}.

However, compliance overhead has emerged as a significant challenge. Implementing conformity assessments and post-market monitoring required additional staffing and expertise, particularly in the area of algorithm auditing. Ensuring regulatory adherence required substantial organizational resources \cite{Schroeder2024}.

A lesson from this experience was that while the structured oversight mechanisms of the EU framework helped strengthen trust and consistency, they also introduced a considerable regulatory burden on organizations. This highlighted the resource-intensive nature of ensuring compliance with high-risk AI regulations.

\subsection{United States: Autonomous Vehicles}
A leading automotive manufacturer piloted a fleet of Level 4 self-driving cars in California. Because the U.S. relies on agency-specific rules, oversight came primarily from the National Highway Traffic Safety Administration (NHTSA) and the state’s Department of Motor Vehicles \cite{Covington2025, CaliforniaDMV2025}.

One major challenge was the presence of fragmented rules across different states. Licensing requirements for driverless vehicles varied significantly, with California mandating proof of safety testing and insurance, while neighboring states imposed fewer constraints \cite{CaliforniaDMV2025,Kirkham2025}. This lack of uniformity created regulatory uncertainty for manufacturers and operators.

Another key issue was public safety and liability. Debates arose over whether manufacturers, software developers, or vehicle operators should bear responsibility in the event of an accident. These discussions highlighted gaps in existing legal frameworks, making it unclear how liability should be assigned \cite{Jansma2016}.

Data governance also posed a challenge, as no single federal law comprehensively regulated the collection of sensor data, such as LiDAR and camera feeds. The absence of clear legal guidance raised significant privacy concerns regarding how such data should be stored, shared, and protected \cite{FTC2024}.

The decentralized regulatory landscape enabled rapid innovation, allowing for early pilot deployments that accelerated technological progress in autonomous vehicles \cite{morley2020initial}. This flexibility encouraged experimentation and advancements in the field.

However, regulatory inconsistency emerged as a significant challenge. The presence of patchwork rules and the absence of uniform standards created uncertainty, making multi-state operations more complex. Additionally, the lack of standardized regulations hindered data-sharing efforts that could have improved safety outcomes.

While the U.S. regulatory model promotes adaptability and innovation, it also risks under-regulating certain high-risk aspects of autonomous vehicle deployment. This underscores the need for more cohesive federal guidelines to balance technological progress with safety and accountability \cite{Kirkham2025, AVIA2025}.

\subsection{United Kingdom: Fintech Regulation}
A UK-based fintech start-up employed AI algorithms to evaluate consumer creditworthiness using non-traditional data sources (e.g., social media, phone usage). Regulators of interest included the Financial Conduct Authority (FCA) and the Information Commissioner’s Office (ICO).

One major challenge was ethical and transparency issues, as the algorithms risked discriminating against certain demographic groups with limited digital footprints. To address this, the Information Commissioner's Office (ICO) recommended meaningful disclosure of data usage to ensure fairness and accountability \cite{ICO2023}.

Another challenge involved sector-specific rules, particularly in consumer finance. The Financial Conduct Authority (FCA) required fintech companies to conduct risk assessments and ongoing audits, including stress-testing AI models under various economic scenarios \cite{FCA2024}. These requirements added layers of compliance but were essential for financial stability and consumer protection.

The issue of proportionate regulation also emerged as a central tension, as startups needed to innovate rapidly while still adhering to consumer protection standards. Striking the right balance between these priorities remained a regulatory challenge \cite{FCA2024}.

Regulatory engagement played a key role in compliance, as early consultations with the FCA and ICO helped guide the development of robust internal compliance checks, fostering consumer trust and regulatory alignment \cite{FCA2023}.

Adaptive oversight allowed the fintech company to scale its operations without hindering product launches. Regulators updated their guidelines incrementally, ensuring that compliance requirements evolved alongside technological advancements \cite{Fintech2023}.

The UK’s flexible sector-by-sector approach can effectively balance innovation and accountability. However, the presence of overlapping guidelines from multiple regulators sometimes created confusion, highlighting the need for clearer coordination between regulatory bodies.

\subsection{China: Facial Recognition in Public Spaces}
A municipal government in China partnered with AI vendors to deploy an extensive facial recognition system for security and public services, including traffic management and criminal identification \cite{Brown2021Public, xu2018chinese}.

One significant challenge was data security and privacy, as the implementation of safeguards under the Data Security Law and the Personal Information Protection Law varied across local agencies \cite{Tan2022}. This inconsistency raised concerns about enforcement and compliance in different jurisdictions.

Another challenge was algorithmic oversight, requiring providers to submit technical specifications to government authorities. However, civil society had limited avenues to scrutinize or challenge these systems, leading to concerns about accountability and transparency \cite{arcesati2021lofty}.

The societal impact of these technologies also became a major issue, particularly regarding continuous surveillance and the potential misuse of collected biometric data \cite{arcesati2021lofty}. Public discourse increasingly focused on the balance between security measures and individual rights.

A notable outcome was the swift rollout of AI-driven facial recognition systems, as centralized directives enabled rapid deployment across multiple city districts. This efficiency demonstrated the advantages of a coordinated, top-down approach to technology adoption.

However, limited transparency was a major drawback, as end-users received minimal information about how facial recognition data was stored or for how long. This lack of disclosure reduced public awareness of potential risks and privacy implications.

While state-led oversight enabled rapid implementation, it also constrained external accountability and public transparency. This case exemplified the ongoing tensions between national security objectives and individual privacy rights, highlighting the need for greater public scrutiny and oversight mechanisms.

\subsection{Synthesis of Case Findings}
These case studies demonstrate the implementation of regulatory principles in practical contexts.

In the EU, structured and risk-based frameworks significantly improve trust and accountability. However, they may present a considerable burden for small and medium-sized enterprises (SMEs) and institutions that do not possess adequate compliance resources.

The US uses a decentralized governance model which encourages rapid pilot projects and encourages innovation. This flexibility helps technological advancements but harms coherent and uniform oversight across various jurisdictions.

The United Kingdom uses a sector-specific regulatory framework, facilitating prompt adjustments to new advancements in artificial intelligence. The participation of various regulators can lead to confusion, especially in intricate AI implementations where multiple regulations may be relevant.

Centralized governance in China allows the rapid and extensive deployment of AI technologies. This approach facilitates rapid deployment; however, it constrains civil society engagement and diminishes public transparency in decision-making processes.

\section{Discussion}\label{sec:discussion}

\subsection{Synthesis of Findings}
Drawing together insights from the comparative analysis and case studies, several critical patterns emerge regarding AI risk management across the European Union (EU), the United States (U.S.), the United Kingdom (UK), and China:

\paragraph{Risk Categorization \& Mitigation} 
Different jurisdictions approach risk management in distinct ways. The EU employs a structured, tiered system, while the U.S. follows a decentralized model, the UK enforces sector-specific rules, and China implements state-centric directives. Although each approach aligns with local legal and cultural norms, all face challenges in keeping pace with the rapid evolution of AI technologies.  

\paragraph{Compliance Requirements \& Burden} 
Compliance obligations vary significantly across regulatory frameworks. The EU and China impose more formal regulatory requirements, whereas the U.S. and UK rely on fragmented or flexible guidelines. This contrast highlights the ongoing trade-off between ensuring comprehensive oversight and fostering innovation.  

\paragraph{Governance \& Oversight} 
Regulatory governance involves multiple stakeholders in all four jurisdictions, including government agencies, industry actors, and, to varying degrees, civil society organizations. However, practical enforcement remains inconsistent, influenced by factors such as agency expertise, available resources, and political will.  

\paragraph{Transparency \& Explainability} 
Different regulatory models emphasize transparency and explainability to varying extents. The EU sets a high standard through conformity assessments and disclosure obligations. In contrast, the U.S. primarily relies on market-driven or sector-specific transparency efforts, the UK adopts a guidance-based approach, and China emphasizes internal audits aligned with national priorities.  

\paragraph{Adaptability \& Innovation} 
Regulatory flexibility and adaptability also differ among jurisdictions. The U.S. and UK prioritize flexible regulatory frameworks, which often come at the cost of consistent oversight. Meanwhile, the EU and China update or extend regulations through formal processes, ensuring stronger controls but potentially slowing adaptation to emerging AI advancements.  

Case studies reinforce these findings, revealing how high-risk domains (e.g., healthcare, autonomous vehicles, fintech, and facial recognition) test each framework’s ability to balance ethical, security, and societal safeguards against the demands of rapid AI development.

\subsection{Implications for Global AI Governance}
Given the diversity of approaches, full harmonization of AI standards remains unlikely in the near future. Divergent policy priorities like protecting civil liberties in the EU and U.S., encouraging sector-led innovation in the UK, and emphasizing social stability in China—create regulatory philosophies that can conflict. However, there is growing international interest in aligning on certain principles, particularly in areas such as high-risk applications, where countries may converge on minimum standards for safety-critical systems, potentially formalized through multinational bodies such as the OECD or ISO. Similarly, data privacy and security remain critical concerns, as cross-border data flows necessitate at least partial interoperability among regulations despite variations in cultural and legal contexts. Additionally, bias and discrimination in AI ethics have garnered increasing attention, with ethical guidelines emphasizing fairness and inclusivity, thereby offering a common foundation for collaborative efforts.

At the same time, frictions remain, particularly around issues of national sovereignty, intellectual property, and the role of state versus private actors in AI governance. Balancing these concerns will require diplomatic engagement, mutual recognition of standards, and ongoing dialogue between global players.

\subsection{Future Directions}
Several key areas warrant further exploration. First, adaptive regulation remains crucial, particularly in light of emergent AI trends such as generative models. The development of legal instruments that can be updated swiftly, without necessitating major legislative overhauls, will be essential in maintaining regulatory relevance.

Second, stakeholder engagement should be expanded to include more systematic involvement of civil society, academic researchers, and underrepresented communities. Such inclusivity can enrich policy debates and help ensure that AI systems align with broader societal values.

Third, international collaboration offers a pathway for governments, international organizations, and industry consortia to coordinate on shared principles, such as risk transparency and algorithmic fairness. While complete harmonization may not be immediately feasible, establishing common guidelines can enhance global AI governance.

Finally, context-specific case studies present an opportunity for future empirical research. Examining additional industries, such as education and agriculture, or focusing on granular local contexts can provide insights into how regulatory frameworks are adapted or circumvented in practice.

As AI technology continues its fast evolution, governance mechanisms must evolve together. A balance of flexibility and rigor, informed by diverse stakeholder input and international cooperation, is likely to make the most responsible and sustainable path forward. 

\section{Conclusion}\label{sec:conclusion}

\subsection{Summary of Main Contributions}
\begin{figure}[htbp]
    \centering
    \resizebox{0.4\textwidth}{!}{\includegraphics{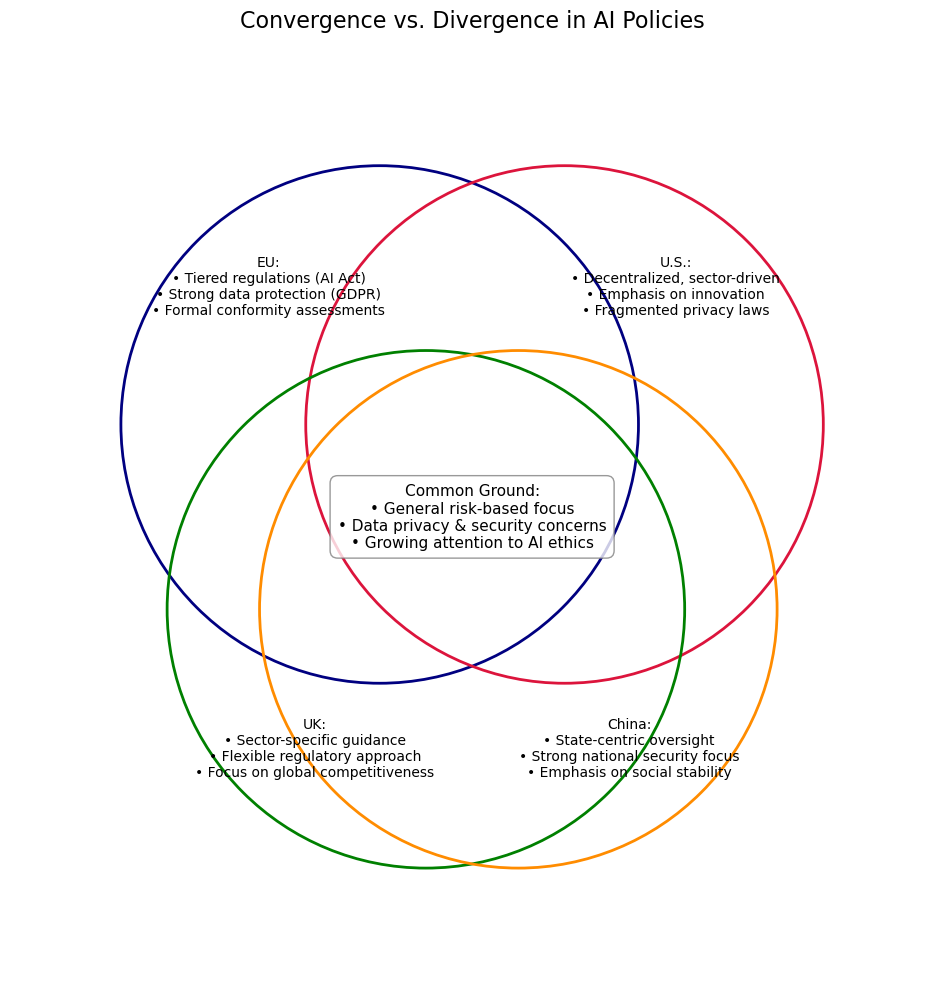}} % Adjust 0.8 for size
    \caption{Convergence vs. Divergence in AI Policies}
    \label{fig:overlapping}
\end{figure}

This paper presents a comparative analysis of AI risk management frameworks across the European Union (EU), United States (U.S.), United Kingdom (UK), and China. It investigates the categorization and mitigation of AI risks, compliance enforcement, governance oversight, transparency promotion, and the balance between adaptability and innovation in each region. The research integrates case studies from high-risk sectors to highlight the real-world challenges and outcomes associated with various regulatory models. This approach addresses our focus research questions:   
\begin{itemize}
    \item \emph{How do these frameworks differ in effectiveness and approach to AI risk mitigation?}  
    \item \emph{What criteria should guide the evaluation of AI governance, and how can frameworks adapt to emerging technologies?}
\end{itemize}
Conclusions include the EU’s structured yet potentially challenging approach, the U.S.’s decentralized but flexible system, the UK’s sector-specific flexibility that can yield inconsistencies, and China’s centralized control which expedites implementation but limits public transparency. These findings reinforce the need for context-sensitive governance strategies that evolve in tandem with rapidly changing AI technologies. \autoref{fig:overlapping} summaries where jurisdictions have overlapping principles and where they diverge.

\subsection{Policy Recommendations}
Several recommendations for regulators, industry stakeholders, and civil society are derived from the comparison analysis.:
An effective AI governance framework should incorporate several key elements. One fundamental aspect is the need to establish clear, proportional standards that differentiate enforcement intensity based on risk levels. By refining tiered regulatory frameworks, policymakers can ensure that low-risk AI applications face minimal compliance burdens while maintaining stringent oversight for high-risk and critical use cases.

Another priority is to promote transparency and explainability in AI systems. Regulators and industry consortia should formulate practical guidelines that facilitate the development of explainable AI models, allowing end-users and impacted communities to better understand and trust AI-driven decisions.

Equally important is the need to encourage multistakeholder engagement. Involving government bodies, technical experts, advocacy groups, and affected communities in policymaking processes ensures that diverse perspectives are considered, leading to more equitable and inclusive AI governance.

Furthermore, governments must invest in adaptive regulatory mechanisms that enable rapid adjustments to emerging AI risks. Legislative instruments and delegated acts should be designed with flexibility to accommodate unforeseen challenges posed by frontier AI technologies, such as generative models.

Lastly, fostering global coordination is essential for harmonizing AI governance across jurisdictions. International organizations and bilateral agreements can play a pivotal role in developing shared ethical standards, facilitating data-sharing protocols, and establishing baseline safety requirements, particularly for AI systems that operate across national borders.

\subsection{Final Reflections}
Effective AI risk management stands at the intersection of technological potential and societal well-being. As AI systems increasingly influence public services, critical infrastructure, and personal decision-making, robust yet flexible governance structures become paramount to harness innovation responsibly. The comparative insights offered here underscore the importance of striking a balance: fostering economic growth and competitive advantages while safeguarding ethical principles, individual rights, and public trust. In a rapidly evolving landscape, policymakers, industry leaders, and civil society must remain vigilant, adapting frameworks to meet new challenges as AI reshapes our world.

\bibliographystyle{unsrt}  
\bibliography{main}  %%% Remove comment to use the external .bib file (using bibtex).

\begin{thebibliography}{100}

\bibitem{bostrom2014paths}
N~Superintelligence Bostrom.
\newblock Paths, dangers, strategies, 2014.

\bibitem{floridi2022unified}
Luciano Floridi and Josh Cowls.
\newblock A unified framework of five principles for ai in society.
\newblock {\em Machine learning and the city: Applications in architecture and urban design}, pages 535--545, 2022.

\bibitem{RASHID2024100277}
Adib~Bin Rashid and MD~Ashfakul~Karim Kausik.
\newblock Ai revolutionizing industries worldwide: A comprehensive overview of its diverse applications.
\newblock {\em Hybrid Advances}, 7:100277, 2024.

\bibitem{dignum2018ethics}
Virginia Dignum.
\newblock Ethics in artificial intelligence: introduction to the special issue.
\newblock {\em Ethics and Information Technology}, 20(1):1--3, 2018.

\bibitem{blobel2020autonomous}
Bernd Blobel, Pekka Ruotsalainen, Mathias Brochhausen, Frank Oemig, and Gustavo~A Uribe.
\newblock Autonomous systems and artificial intelligence in healthcare transformation to 5p medicine--ethical challenges.
\newblock In {\em Digital personalized health and medicine}, pages 1089--1093. IOS Press, 2020.

\bibitem{Pratt2022}
B~Pratt, M~Parker, and S~Bull.
\newblock Equitable design and use of digital surveillance technologies during covid-19: Norms and concerns.
\newblock {\em J Empir Res Hum Res Ethics}, 17(5):573--586, Dec 2022.
\newblock Epub 2022 Sep 7.

\bibitem{batool2023responsible}
Amna Batool, Didar Zowghi, and Muneera Bano.
\newblock Responsible ai governance: a systematic literature review.
\newblock {\em arXiv preprint arXiv:2401.10896}, 2023.

\bibitem{eu2021proposal}
EU~Commission et~al.
\newblock Proposal for a regulation laying down harmonised rules on artificial intelligence.
\newblock {\em Brussels}, 21:2021, 2021.

\bibitem{Almada2025}
Marco Almada.
\newblock The eu ai act in a global perspective.
\newblock In J~Furendal and B~Lundgren, editors, {\em Handbook on the Global Governance of AI}. Edward Elgar, 2025.
\newblock Forthcoming.

\bibitem{cath2018artificial}
Corinne Cath, Sandra Wachter, Brent Mittelstadt, Mariarosaria Taddeo, and Luciano Floridi.
\newblock Artificial intelligence and the ‘good society’: the us, eu, and uk approach.
\newblock {\em Science and engineering ethics}, 24:505--528, 2018.

\bibitem{morley2023operationalising}
Jessica Morley, Libby Kinsey, Anat Elhalal, Francesca Garcia, Marta Ziosi, and Luciano Floridi.
\newblock Operationalising ai ethics: barriers, enablers and next steps.
\newblock {\em AI \& SOCIETY}, pages 1--13, 2023.

\bibitem{lords2018ai}
House~Of Lords et~al.
\newblock Ai in the uk: ready, willing and able?
\newblock {\em Retrieved August}, 13:2021, 2018.

\bibitem{ding2018deciphering}
Jeffrey Ding.
\newblock Deciphering china’s ai dream.
\newblock {\em Future of Humanity Institute Technical Report}, 2018.

\bibitem{Calero2024}
Hipolito Calero.
\newblock An analysis of china’s ai governance proposals, September 12 2024.
\newblock Blog post.

\bibitem{eubanks2018automating}
Virginia Eubanks.
\newblock {\em Automating inequality: How high-tech tools profile, police, and punish the poor}.
\newblock St. Martin's Press, 2018.

\bibitem{10.1093/polsoc/puae020}
Brian Judge, Mark Nitzberg, and Stuart Russell.
\newblock When code isn’t law: rethinking regulation for artificial intelligence.
\newblock {\em Policy and Society}, page puae020, 05 2024.

\bibitem{bryan2024balancing}
Kevin~A Bryan and Florenta Teodoridis.
\newblock Balancing market innovation incentives and regulation in ai: Challenges and opportunities.
\newblock {\em The Brookings Institution}, 2024.

\bibitem{Li2024}
Cathy Li.
\newblock Balancing innovation and governance in the age of ai.
\newblock {\em World Economic Forum}, November 2024.

\bibitem{chen2023ethics}
Zhisheng Chen.
\newblock Ethics and discrimination in artificial intelligence-enabled recruitment practices.
\newblock {\em Humanities and Social Sciences Communications}, 10(1):1--12, 2023.

\bibitem{TurnerLee2022}
Nicol~Turner Lee and Caitlin Chin.
\newblock Police surveillance and facial recognition: Why data privacy is imperative for communities of color.
\newblock {\em Brookings Institution}, April 2022.

\bibitem{DHS2024}
{U.S. Department of Homeland Security}.
\newblock Groundbreaking framework for the safe and secure deployment of ai in critical infrastructure, November 2024.

\bibitem{voigt2017eu}
Paul Voigt and Axel Von~dem Bussche.
\newblock The eu general data protection regulation (gdpr).
\newblock {\em A Practical Guide, 1st Ed., Cham: Springer International Publishing}, 10(3152676):10--5555, 2017.

\bibitem{edwards2021eu}
Lilian Edwards.
\newblock The eu ai act: a summary of its significance and scope.
\newblock {\em Artificial Intelligence (the EU AI Act)}, 1, 2021.

\bibitem{OECD2023}
{OECD}.
\newblock The state of implementation of the oecd ai principles four years on.
\newblock Technical Report~3, OECD Publishing, Paris, 2023.

\bibitem{liebl2023ai}
Andreas Liebl and Till Klein.
\newblock Ai act: Risk classification of ai systems from a practical perspective.
\newblock Technical report, Retrieved 2024-03-11 from https://www. appliedai. de/assets/files/AI~…, 2023.

\bibitem{schuett2024risk}
Jonas Schuett.
\newblock Risk management in the artificial intelligence act.
\newblock {\em European Journal of Risk Regulation}, 15(2):367--385, 2024.

\bibitem{Demkova2025}
Simona Demková and Giovanni~De Gregorio.
\newblock The looming enforcement crisis in european digital policy: A rule-of-law centered path forward.
\newblock {\em VerfBlog}, February 2025.

\bibitem{bateman2022us}
Jon Bateman.
\newblock Us-china technological “decoupling”: A strategy and policy framework, 2022.

\bibitem{ai2023artificial}
NIST AI.
\newblock Artificial intelligence risk management framework (ai rmf 1.0), 2023.

\bibitem{OSTP2022}
{White House Office of Science and Technology Policy}.
\newblock Blueprint for an ai bill of rights, October 2022.

\bibitem{2TurnerLee2022}
Nicol~Turner Lee and Jack Malamud.
\newblock Opportunities and blind spots in the white house's blueprint for an ai bill of rights.
\newblock {\em Brookings Institution}, December 2022.

\bibitem{UKGovAIRegulation2024}
{Department for Science, Innovation \& Technology}.
\newblock A pro-innovation approach to ai regulation: Government response, February 2024.

\bibitem{singh2024artificial}
Charanjit Singh.
\newblock Artificial intelligence and deep learning: considerations for financial institutions for compliance with the regulatory burden in the united kingdom.
\newblock {\em Journal of Financial Crime}, 31(2):259--266, 2024.

\bibitem{Hickman2025}
Tim Hickman, Jenna Rennie, and Aishwarya Jha.
\newblock Ai watch: Global regulatory tracker - united kingdom.
\newblock {\em White \& Case LLP}, February 2025.

\bibitem{Davies2023}
Matt Davies and Michael Birtwistle.
\newblock Regulating ai in the uk: Strengthening the uk's proposals for the benefit of people and society, July 2023.

\bibitem{HouseOfLordsLibrary2023}
{House of Lords Library}.
\newblock Artificial intelligence: Development, risks and regulation, July 2023.

\bibitem{Wang2024}
Wang Yi.
\newblock Promoting development for all and bridging the ai divide, September 2024.

\bibitem{cheng2023shaping}
Jing Cheng and Jinghan Zeng.
\newblock Shaping ai’s future? china in global ai governance.
\newblock {\em Journal of Contemporary China}, 32(143):794--810, 2023.

\bibitem{chen2024developing}
Meng Chen.
\newblock Developing china's approaches to regulate cross-border data transfer: Relaxation and integration.
\newblock {\em Computer Law \& Security Review}, 54:105997, 2024.

\bibitem{creemers2021china}
Rogier Creemers.
\newblock China’s emerging data protection framework.
\newblock {\em Journal of Cybersecurity}, 8(1):tyac011, 08 2022.

\bibitem{ye2024privacy}
Xiongbiao Ye, Yuhong Yan, Jia Li, and Bo~Jiang.
\newblock Privacy and personal data risk governance for generative artificial intelligence: A chinese perspective.
\newblock {\em Telecommunications Policy}, 48(10):102851, 2024.

\bibitem{luna2024navigating}
Jose Luna, Ivan Tan, Xiaofei Xie, and Lingxiao Jiang.
\newblock Navigating governance paradigms: A cross-regional comparative study of generative ai governance processes \& principles.
\newblock In {\em Proceedings of the AAAI/ACM Conference on AI, Ethics, and Society}, volume~7, pages 917--931, 2024.

\bibitem{schmitz2024global}
Anna Schmitz, Michael Mock, Rebekka G{\"o}rge, Armin~B Cremers, and Maximilian Poretschkin.
\newblock A global scale comparison of risk aggregation in ai assessment frameworks.
\newblock {\em AI and Ethics}, pages 1--26, 2024.

\bibitem{batool2025ai}
Amna Batool, Didar Zowghi, and Muneera Bano.
\newblock Ai governance: a systematic literature review.
\newblock {\em AI and Ethics}, pages 1--15, 2025.

\bibitem{hantrais2008international}
Linda Hantrais.
\newblock {\em International comparative research: Theory, methods and practice}.
\newblock Bloomsbury Publishing, 2008.

\bibitem{braun2006using}
Virginia Braun and Victoria Clarke.
\newblock Using thematic analysis in psychology.
\newblock {\em Qualitative research in psychology}, 3(2):77--101, 2006.

\bibitem{stake1995case}
Robert Stake.
\newblock {\em Case study research}.
\newblock Springer, 1995.

\bibitem{creswell2016qualitative}
John~W Creswell and Cheryl~N Poth.
\newblock {\em Qualitative inquiry and research design: Choosing among five approaches}.
\newblock Sage publications, 2016.

\bibitem{wu2023comprehensive}
Weiyue Wu and Shaoshan Liu.
\newblock A comprehensive review and systematic analysis of artificial intelligence regulation policies.
\newblock {\em arXiv preprint arXiv:2307.12218}, 2023.

\bibitem{lost_in_translation}
Sabine Mokry.
\newblock What is lost in translation? differences between chinese foreign policy statements and their official english translations.
\newblock {\em Foreign Policy Analysis}, 18(3):orac012, 04 2022.

\bibitem{EUAIArticle72}
{European Union}.
\newblock Article 72: Post-market monitoring by providers and post-market monitoring plan for high-risk ai systems, 2024.

\bibitem{Lumenova2024}
{Lumenova AI}.
\newblock Decoding the eu ai act: Transparency and governance, March 2024.

\bibitem{Zaidan2024}
E.~Zaidan and I.~A. Ibrahim.
\newblock Ai governance in a complex and rapidly changing regulatory landscape: A global perspective.
\newblock {\em Humanities and Social Sciences Communications}, 11:1121, 2024.

\bibitem{AlgorithmicAccountabilityAct2023}
U.S. Senate.
\newblock Algorithmic accountability act of 2023, s.2892, 118th congress, September 2023.

\bibitem{Lively2021}
Taylor~Kay Lively.
\newblock Facial recognition in the united states: Privacy concerns and legal developments.
\newblock {\em Security Management}, December 2021.

\bibitem{DataProtectionAct2018}
{Parliament of the United Kingdom}.
\newblock Data protection act 2018, 2018.

\bibitem{ChinaSocialCredit}
Genia Kostka.
\newblock China’s social credit systems and public opinion: Explaining high levels of approval.
\newblock {\em New Media \& Society}, 21(7):1565--1593, 2019.

\bibitem{LathamWatkins2023}
{Latham \& Watkins Privacy \& Cyber Practice}.
\newblock China’s new ai regulations, August 2023.

\bibitem{ForeignPolicy2022}
{Foreign Policy}.
\newblock Why china's new data security law is a warning for the future of data governance, January 2022.

\bibitem{Zenner2024}
Kai Zenner.
\newblock The ai act: Responsibilities of the eu member states, August 2024.

\bibitem{TimeThierryBreton2024}
TIME.
\newblock Thierry breton.
\newblock {\em TIME}, September 2024.

\bibitem{EPLegislativeTrain2024}
{European Parliament}.
\newblock Artificial intelligence act: Legislative train schedule, December 2024.

\bibitem{smith2018challenges}
Melanie Smith.
\newblock Challenges in the implementation of eu law at national level, 2018.

\bibitem{Demirci2024}
Suleyman Demirci.
\newblock Empowering small businesses: The impact of ai on leveling the playing field, March 2024.

\bibitem{OpenSecrets2024}
{OpenSecrets}.
\newblock Federal lobbying on artificial intelligence grows as legislative efforts stall, January 2024.

\bibitem{morley2020initial}
Jessica Morley, Luciano Floridi, Libby Kinsey, and Anat Elhalal.
\newblock From what to how: an initial review of publicly available ai ethics tools, methods and research to translate principles into practices.
\newblock {\em Science and engineering ethics}, 26(4):2141--2168, 2020.

\bibitem{WhittakerSmith2024}
Tom Whittaker and Liz Smith.
\newblock Ai law, regulation and policy - highlights from 2024 and what to look forward to in 2025, December 2024.

\bibitem{CDEI2025}
{Department for Science, Innovation and Technology}.
\newblock Centre for data ethics and innovation (cdei), 2025.

\bibitem{Kaleka2024}
Dharminder~Singh Kaleka.
\newblock The ai gambit: Will the uk lead or follow?, October 2024.

\bibitem{LOC2023}
{Laney Zhang and Library of Congress}.
\newblock China: Generative ai measures finalized, July 2023.

\bibitem{ICNL2024}
{International Center for Not-for-Profit Law}.
\newblock China civic freedom monitor, December 2024.

\bibitem{F52024}
{Lori MacVittie}.
\newblock Crucial concepts in ai: Transparency and explainability, July 2024.

\bibitem{gdpr2022automated}
EU~GDPR.
\newblock Automated individual decision-making, including profiling, 2022.

\bibitem{Schwaeke13082024}
Julia Schwaeke, Anna Peters, Dominik~K. Kanbach, Sascha Kraus, and Paul Jones.
\newblock The new normal: The status quo of ai adoption in smes.
\newblock {\em Journal of Small Business Management}, 0(0):1--35, 2024.

\bibitem{schmude2025information}
Timoth{\'e}e Schmude, Laura Koesten, Torsten M{\"o}ller, and Sebastian Tschiatschek.
\newblock Information that matters: Exploring information needs of people affected by algorithmic decisions.
\newblock {\em International Journal of Human-Computer Studies}, 193:103380, 2025.

\bibitem{FDA2025}
{U.S. Food and Drug Administration}.
\newblock Fda issues comprehensive draft guidance for developers of artificial intelligence-enabled medical devices, January 2025.

\bibitem{FTC2025}
Staff in~the Office~of Technology and the Division~of Advertising~Practices.
\newblock Ai and the risk of consumer harm, January 2025.

\bibitem{Jones2021}
Nigel Jones.
\newblock 10 reasons to be concerned about facial recognition technology, August 2021.

\bibitem{Guardian2024}
Johana Bhuiyan.
\newblock She didn't get an apartment because of an ai-generated score – and sued to help others avoid the same fate.
\newblock {\em The Guardian}, December 2024.

\bibitem{ICO2025}
{Information Commissioner's Office (ICO)}.
\newblock Rights related to automated decision-making, including profiling, 2025.

\bibitem{GLI_BankingUK2024}
{ Alastair Holt and Simon Treacy}.
\newblock Banking \& finance laws and regulations – united kingdom, 2024.

\bibitem{roberts2023artificial}
Huw Roberts, Alexander Babuta, Jessica Morley, Christopher Thomas, Mariarosaria Taddeo, and Luciano Floridi.
\newblock Artificial intelligence regulation in the united kingdom: a path to good governance and global leadership?
\newblock {\em Internet Policy Review}, 2023.

\bibitem{ChinaAIInterim2023}
{China Law Translate}.
\newblock Interim measures for the management of generative artificial intelligence services, July 2023.

\bibitem{menglu2024regulation}
Wang Menglu.
\newblock Regulation of algorithmic decision-making in china: Development, problems and implications.
\newblock {\em Singapore Journal of Legal Studies}, 1:276--305, 2024.

\bibitem{llorca2024testing}
David~Fern{\'a}ndez Llorca, Ronan Hamon, Henrik Junklewitz, Kathrin Grosse, Lars Kunze, Patrick Seiniger, Robert Swaim, Nick Reed, Alexandre Alahi, Emilia G{\'o}mez, et~al.
\newblock Testing autonomous vehicles and ai: perspectives and challenges from cybersecurity, transparency, robustness and fairness.
\newblock {\em arXiv preprint arXiv:2403.14641}, 2024.

\bibitem{DigitalRegulation_AI_Challenges_2024}
{Digital Regulation Platform}.
\newblock Transformative technologies (ai) challenges and principles of regulation, May 2024.

\bibitem{ITIF2021}
{Information Technology and Innovation Foundation (ITIF)}.
\newblock How much will the artificial intelligence act cost europe?, July 2021.

\bibitem{Orrick2024}
Julia Apostle and Haley Flora.
\newblock The eu ai act: 10 things startups should know, October 2024.

\bibitem{Bhatti2024}
Ayesha Bhatti.
\newblock An agile, sector-specific approach to uk ai regulation is promising, 2024.

\bibitem{zhang2024promise}
Angela~Huyue Zhang.
\newblock The promise and perils of china's regulation of artificial intelligence.
\newblock {\em Available at SSRN}, 2024.

\bibitem{deepc2023}
deepc.
\newblock Artificial intelligence for better diagnostics: Lmu university hospital munich chooses deepc as ai platform partner, 2023.

\bibitem{James2024}
Ted~A. James.
\newblock Confronting the mirror: Reflecting on our biases through ai in health care.
\newblock {\em Trends in Medicine}, September 2024.

\bibitem{healthCareServices}
Janos Meszaros, Jusaku Minari, and Isabelle Huys.
\newblock The future regulation of artificial intelligence systems in healthcare services and medical research in the european union.
\newblock {\em Frontiers in Genetics}, 13, 2022.

\bibitem{sartor2020impact}
Giovanni Sartor, Francesca Lagioia, et~al.
\newblock The impact of the general data protection regulation (gdpr) on artificial intelligence.
\newblock {\em European Parliamentary Research Service}, 2020.

\bibitem{Jeong2025}
Jinseo Jeong, Sohyun Kim, Lian Pan, Daye Hwang, Dongseop Kim, Jeongwon Choi, Yeongkyo Kwon, Pyeongro Yi, Jisoo Jeong, and Seok-Ju Yoo.
\newblock Reducing the workload of medical diagnosis through artificial intelligence: A narrative review.
\newblock {\em Medicine (Baltimore)}, 104(6):e41470, February 2025.

\bibitem{Schroeder2024}
Pascal~Yves Schroeder, Catharina Glugla, Alex Shandro, Catherine~Di Lorenzo, Sarah~De Wulf, Filip~Van Elsen, David Wakeling, Jane Finlayson-Brown, Justyna Ostrowska, Laur Badin, Laurie-Anne Ancenys, Nicole~Wolters Ruckert, Paul Wagner, Peter~Van Dyck, Robert Dickens, Ross Phillipson, and Steve Wood.
\newblock Zooming in on ai – \#10: Eu ai act – what are the obligations for high-risk ai systems?, October 2024.

\bibitem{Covington2025}
Covington \&~Burling LLP.
\newblock Nhtsa proposes new autonomous vehicle program, January 2025.

\bibitem{CaliforniaDMV2025}
{California Department of Motor Vehicles (DMV)}.
\newblock California autonomous vehicle regulations, 2025.

\bibitem{Kirkham2025}
Chris Kirkham and Abhirup Roy.
\newblock Tesla robotaxis by june? musk turns to texas for hands-off regulation.
\newblock {\em Reuters}, February 2025.

\bibitem{Jansma2016}
Steven~D. Jansma.
\newblock Autonomous vehicles: The legal landscape in the us, August 2016.

\bibitem{FTC2024}
{Federal Trade Commission (FTC)}.
\newblock Cars and consumer data: Unlawful collection and use, May 2024.

\bibitem{AVIA2025}
{Autonomous Vehicle Industry Association (AVIA)}.
\newblock Autonomous vehicle industry association unveils federal policy framework to advance safe deployment of avs, January 2025.

\bibitem{ICO2023}
{Information Commissioner's Office (ICO)}.
\newblock Guidance on ai and data protection, March 2023.

\bibitem{FCA2024}
{Financial Conduct Authority (FCA)}.
\newblock Ai update, April 2024.

\bibitem{FCA2023}
{Financial Conduct Authority (FCA)}.
\newblock Innovation: engagement, May 2023.

\bibitem{Fintech2023}
Martin Cook and Matthew Loader.
\newblock Fintech: Fca and ico joint letter provides clarity for firms navigating the interplay between consumer duty and data protection requirements, July 2023.

\bibitem{Brown2021Public}
Tristan~G. Brown, Alexander Statman, and Celine Sui.
\newblock Public {Debate} on {Facial} {Recognition} {Technologies} in {China}.
\newblock {\em MIT Case Studies in Social and Ethical Responsibilities of Computing}, (Summer 2021), aug 10 2021.
\newblock https://mit-serc.pubpub.org/pub/public-debate-on-facial-recognition-technologies-in-china.

\bibitem{xu2018chinese}
Vicky~Xiuzhong Xu and Bang Xiao.
\newblock Chinese authorities use facial recognition, public shaming to crack down on jaywalking, criminals.
\newblock {\em ABC News}, 20, 2018.

\bibitem{Tan2022}
Michael Tan, Mike Goldammer, Julian Sun, and Kyle Tong.
\newblock Prc data protection law: How an effective compliance management system may help to reduce liabilities, June 2022.

\bibitem{arcesati2021lofty}
Rebecca Arcesati.
\newblock Lofty principles, conflicting incentives: Ai ethics and governance in china.
\newblock {\em Merics. https://merics. org/en/report/lofty-principles-conflicting-incentives-ai-ethics-andgovernance-china}, 2021.

\end{thebibliography}


\end{document}